\documentstyle[epsf]{l-aa} 

\catcode`\"=\active\let"=\"
\def\3{\ss }
\newcommand{\hd}{HD\,149499}

\newcommand{\hdb}{HD\,149499\,B}
\newcommand{\nhydro}{n_{\rm{H}}/n_{\rm{He}}}
\newcommand{\nhelium}{n_{\rm{He}}/n_{\rm{H}}}
\newcommand{\ion}[2]{#1\,{\sc #2}}
\newcommand{\teff}{T_{\rm eff}}         
\def\fm{\hbox{$.\!\!^{\rm{m}}$}}
\def\mag{\hbox{$\hbox{\thinspace \fm}$}}

\newcommand{\gppr}{\stackrel{>}{\scriptstyle \sim}}
\newcommand{\gappr}{\raisebox{-0.4ex}{$\gppr $}}

\newcommand{\litanf}{\begin{list}{}{\leftmargin=1.5cm \rightmargin=0cm 
\itemindent=-1.5cm \parsep=0cm \itemsep=0cm }}
\newcommand{\litend}{\end{list}}

\newcommand{\apj}{ApJ}

\newcommand{\aua}{A\&A}

\begin{document}

\thesaurus{ 08 (08.06.3; 08.09.2: \hdb; 8.23.1)}

\title{Temperature determination of the cool DO white dwarf \hdb\ from EUVE
observations\thanks{Based on observations with the EUVE and ROSAT satellites}
}
\author{S.~Jordan\inst{1} \and R.~Napiwotzki\inst{2}  \and D.~Koester\inst{1}
	\and T.~Rauch\inst{1}\inst{3}}
\institute{
Institut f\"ur Astronomie und Astrophysik der Universit\"at,
24098 Kiel, Germany
\and
Dr.~Remeis-Sternwarte, Sternwartstr.~7, 96049~Bamberg, Germany
\and
        Lehrstuhl Astrophysik, Universit\"at Potsdam,
        Am Neuen Palais 10, D-14469 Potsdam, Germany
}
 \offprints{S. Jordan}
\date{Received date; accepted date}

\maketitle
\markboth{S.~Jordan et al.: EUVE observations of \hdb}{S.~Jordan et al.: EUVE observations of \hdb}

\begin{abstract}
We present a LTE model atmosphere analysis of the EUVE medium and long
 wavelength spectrum of the cool DO white dwarf \hdb. This observation
 in the spectral range 230-700\,\AA\ supplements a previous analysis
 of an ORFEUS spectrum between 912 and 1170\,\AA\ which yielded the
 atmospheric parameters $\teff=49500\pm 500$\,K and $\log g = 7.97\pm
 0.08$ and a hydrogen-to-helium number ratio of $0.22\pm 0.11$.  The
 EUVE data are in full agreement with the ORFEUS result and allow a
 more precise determination of the effective temperature
 ($\teff=49500\pm 200$\,K) and the interstellar hydrogen column
 density ($N_{\rm H}=7 \cdot 10^{18}$\,cm$^{-2}$).  None of the
 features in the EUVE spectrum could be identified with any additional
 absorber besides helium. Exploratory calculations with fully metal
 blanketed LTE model atmospheres show that the metal abundances
 predicted by current diffusion theory are clearly at variance with
 the observed spectrum.


\end{abstract}

\section{Introduction}
White dwarfs can be divided into two groups: the spectral class DA
 showing only the Balmer lines of hydrogen in their optical spectrum
 and the non-DA white dwarfs. The atmospheres of the latter are mostly
 dominated by helium. Depending on the effective temperature ($T_{\rm
 eff}$), the spectra are either characterized by \ion{He}{ii} (DO
 stars with $T_{\rm eff}\gappr 45\,000$\,K are also showing
\ion{He}{i} lines at the cool end), \ion{He}{i} lines (spectral class
DB with $11\,000$\,K$ < T_{\rm eff} < 28\,000$\,K) or exhibit a
continuous spectrum in the optical ($T_{\rm eff} < 11\,000$\,K).
Between the DO stars and the DB stars a hitherto unexplained gap
exists: no hydrogen poor star is found between 28\,000 and 45\,000\,K
(Liebert et al 1986).  Possibly the non-DA sequence is the progeny of
the PG\,1159 stars which are helium, carbon, and oxygen rich.

Since even the FUV wavelength range of the DO stars ($\teff \geq
50\,000$\,K) is on the Rayleigh-Jeans tail of the electromagnetic
spectrum, no accurate temperature determination can be expected from
continuum fitting.  More promising is the use of line profiles with
elaborate model atmospheres, especially the utilization of ionization
equilibria (e.g.\ \ion{He}{i}/\ion{He}{ii}) by fitting lines of
different ionization stages for the $\teff$ determination; this
approach has e.g. been used by Koester et al.\ (1979) for a LTE
analysis of the cool DO HZ\,21.  The most comprehensive spectral
analysis of DO stars with NLTE model atmospheres was performed by
Dreizler \&\ Werner (1996).  They could not identify hydrogen neither
in optical nor in UV spectra; in the cooler DAs hydrogen should be
easily detectable in the optical if the hydrogen abundances exceed
about 10\%.  Therefore, HD\,149499\,B remains the only star with
strong --- though indirect --- indications for the presence of
hydrogen (Napiwotzki et al. 1995).

With $V \approx 11\mag 7$ HD\,149499\,B, the secondary of a binary, is
$1\mag 5$ brighter than any other DO white dwarf.  The KOV primary ($V
\approx 8\mag 2$) of the system is only $1.5^{\prime \prime}$ away and
strongly contaminates the optical spectrum.

In the UV/FUV region the white dwarf flux dominates at least up to
2100\,\AA\ (Sion et al.\ 1982, Wray et al.\ 1979).  However, since it
is not possible to detect the \ion{He}{i} lines in the optical region,
the temperature determination of this DO remained controversial until
recently: Sion et al.\ (1982) determined
$\teff=55\,000^{\hspace{2mm}+5\,000}_{-15\,000}$\,K from a fit of the
\ion{He}{ii}\ lines in the IUE range and the comparison of the
continuum flux.  In disagreement with this result, Wray et al.\ (1979)
derived $70\,000$\,K$\leq \teff \leq 100\,000$\,K from the continuum
fit only.

Napiwotzki et al.\ (1995) performed an analysis of a FUV spectrum of
\hdb\ taken with the EUV/FUV Berkeley spectrometer of the ORFEUS experiment
and obtained the first reliable temperature determination of \hdb:
 $\teff = 49\,500\pm 500$\,K.  Their most important result was that
 photospheric hydrogen could be inferred from the ORFEUS spectrum.
 The closeness of \hdb\ to the DB gap, in which all white dwarfs
 probably become hydrogen rich makes it a very unique object for the
 understanding of how the transition of former DO white dwarfs into
 DAs below about 50\,000\,K may occur.

Since no \ion{He}{i} line is detectable in the FUV, the temperature
and gravity determination of \hdb\ by Napiwotzki et al.\ (1995) is
based on a simultaneous fit of the \ion{He}{ii} $2\rightarrow n$
(Balmer) series in the ORFEUS wavelength range, which limits the
accuracy of the temperature. Since no broadening tables with up to
date theories are available for these lines, systematic errors cannot
be excluded.  Thus we complement the analysis of \hdb\ with data
gathered with the EUVE satellite in order to put further constraints
on the atmospheric parameters of \hdb.

The EUV is well suited for the detection of metal features in hot
white dwarfs (e.g. Vennes et al.\ 1989, Barstow et al.\ 1995, Jordan
et al.\ 1996a ).  Therefore we tried to identify lines of heavy
elements and calculated model spectra to check the predictions of
diffusion theory. Currently, RE0503-289 is the only DO white dwarf in
which features of carbon and nitrogen could be identified in an EUV
spectrum (Barstow et al. 1995).

\section{Observations}

The EUV spectra of \hd\ were obtained during June 23-27, 1993 with the
EUVE satellite; the effective exposure time was 91732\,sec.  The three
spectrometers of EUVE cover the wavelength intervals $70-190$\,\AA\
(short-wavelength; SW), $140-380$\,\AA\ (medium-wavelength; MW), and
$280-760$\,\AA\ (long-wavelength; LW) with a spectral resolution of
$\lambda/\Delta \lambda \approx 300$.  Since the spectra have not been
observed in dithered mode it is likely that some of the observed
features are caused by fixed pattern noise.

Significant flux was detected in the MW and LW ranges only.  The first
lines of the \ion{He}{ii}\ resonance series (303.8, 256.3, and
243.0\,\AA) are clearly visible (Fig.\,\ref{xmwfit04800} and
\ref{xlwfit04800}). Napiwotzki et al. (1996) performed a preliminary
analysis of the EUVE spectra using the data from the EUVE standard
extraction. However, in Fig.\,2 of their paper a 20\%\ inconsistency
between the flux measured with the MW and LW spectrograph,
respectively, can be noticed. Thus, we have repeated the data
reduction for this paper with the EUV package in IRAF\footnote{IRAF is
distributed by the National Optical Astronomy Observatories, which is
operated by the Association of Universities for Research in Astronomy,
Inc. (AURA) under cooperative agreement with the National Science
Foundation} with the result that the inconsistency disappeared (see
Sect.\,\ref{euveanalysis}).

The \hd\ system was observed with ROSAT during a pointed observation
on March 15, 1993, with an exposure time of 1376 sec.  ROSAT carries
aboard two instruments: a X-ray telescope, which we used in connection
with the PSPC (positional sensitive proportional counter) detector,
and the Wide Field Camera (WFC). The PSPC covers approximately the
energies between 0.1\,keV and 2.4\,keV ($5-100$\,\AA) with a modest
energy resolution.

 The WFC measurement was made through the P1 filter. No flux was
detected in the corresponding spectral range ($150-220$\,\AA). In the
ROSAT WFC All-Sky Survey (Pye et al. 1995), where the S1
($60-140$\,\AA) and S2 ($112-200$\,\AA) filters were used,
\hdb\ also remained undetected. 

\section{Analysis}
\label{analysis}

\subsection{The ROSAT observation}
\begin{figure}[htbp]
\epsfxsize=8.8cm
\epsffile{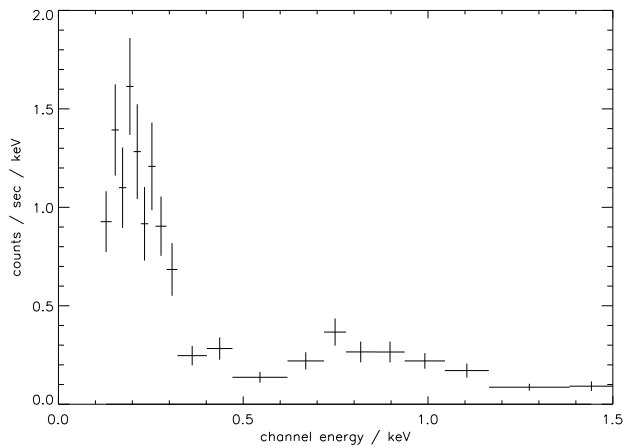}
\caption[]{ROSAT PSPC pulse height distribution of HD\,149499. Since no
significant flux is expected from the DO component \hdb, the X-ray
 radiation must be emitted by the K star primary }
\label{hdrosat}
\end{figure}

Typically, no X-ray flux of DO white dwarfs ist measured above
0.4\,keV (see e.g. Werner et al. 1995). However, the observed pulse
height distribution (PHD, Fig.\,\ref{hdrosat}) of \hdb\ exhibits flux
up to 1.5\,keV. We have reduced and analyzed the ROSAT data as
described in Jordan et al. (1994). The predicted PHD for a model with
the parameters derived from the ORFEUS and EUVE analysis (see
Sect.\,\ref{analysis}) is a factor of $10^{6}$ lower than observed
($0.48\pm 0.02$ counts/sec).  Therefore we conclude that practically
all of the measured soft X-ray flux originates from the K0V star
companion.  We adopted the conversion factor $6\cdot
10^{-12}$\,erg\,cm$^{-2}$\,ct$^{-1}$ (Hempelmann et al. 1995) and the
distance $d=34$\,pc (Ianna et al.\ 1982) and computed an X-ray
luminosity $L_{\rm X} = 4.0\cdot 10^{29}$\,erg/sec. This corresponds
to a relative X-ray luminosity $L_{\rm X}/L_{\rm bol} = 2.5\cdot
10^{-4}$. From a comparison with the ROSAT sample of late type main
sequence stars presented by Hempelmann et al.\ (1995), we conclude
that HD\,149499\,A displays high activity. Further insights may be
gained from a measurement of the rotational period of this star.

\subsection{The EUVE spectra}
\label{euveanalysis}
\subsubsection{Temperature determination}
The LTE model atmosphere codes of Koester were used to calculate an
 extensive grid of H$+$He models in the relevant parameter range of
 \hdb.  The model code uses the classical assumptions of LTE models:
 plane-parallel stratification, hydrostatic and radiative
 equilibrium. Convection is taken into account using the ML1 version
 of the mixing length approximation with l/H$_p$ =1.

With the help of these model atmospheres Napiwotzki et al. (1995) have
analyzed the FUV spectrum of \hdb\ obtained with the Berkeley EUV/FUV
spectrometer of the ORFEUS experiment. Significant flux was detected
above the Lyman edge of hydrogen up to $1170$\,\AA.  The basic
atmospheric parameters were obtained by a simultaneous fit of the
\ion{He}{ii} $2\rightarrow n$ series and the coinciding Lyman lines
yielding $\teff=49500\pm 500$\,K, $\log g = 7.97\pm 0.08$, and
$\nhydro = 0.22\pm 0.11$.

In a first step we tried to reproduce the EUVE spectra with synthetic
spectra calculated with the atmospheric parameters derived from the
ORFEUS analysis. Just as in the case of the ORFEUS FUV spectrum we
normalized the model flux according to the observed flux at 1150\,\AA.
Subsequently, the predicted photon flux was convolved with the
detector response of the spectrograph taking into account higher order
contamination.  The interstellar absorption was calculated according
to Rumph et al. (1994) and Morrison \&\ MacCammon(1983).

\begin{figure}[htbp]
\epsfxsize=8.8cm
\epsffile{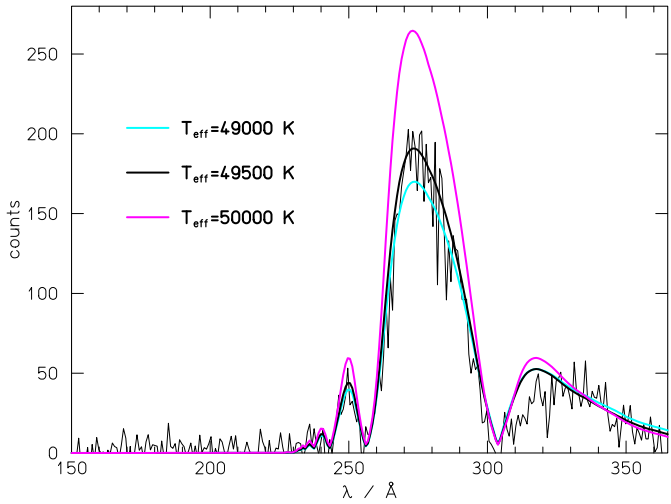}
\caption[]{Comparison of the EUVE medium wavelength spectrum with 
theoretical models with $T_{\rm eff}=49\,000$, 49\,500, and
50\,000\,K. Consistent with the result from the ORFEUS analysis, $\log
g=8$ and a H/He ratio of 0.22 was assumed. For the three different
temperatures hydrogen column densities of $7.1$, $7.6$, and $8.5 \cdot
10^{18}$\,cm$^{-2}$ were used }
\label{xmwfit04800}
\end{figure}

\begin{figure}[htbp]
\epsfxsize=8.8cm
\epsffile{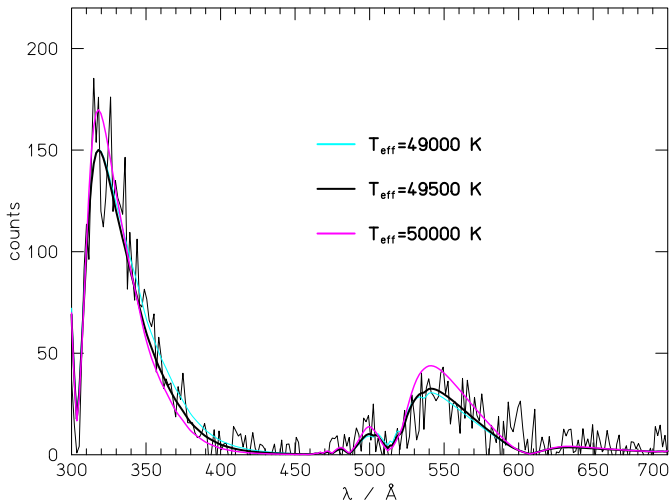}
\caption[]{The long wavelength EUVE spectrum of \hdb\ is compared with
the predictions for model atmospheres having the same atmospheric
parameters as listed in Fig.\,\ref{xmwfit04800}. Note that the
measured count rate above 456\,\AA\ is entirely due to second order
contamination from the flux between 228 and 350\,\AA}
\label{xlwfit04800}
\end{figure}

In Fig.\,\ref{xmwfit04800} and \ref{xlwfit04800} we compare the EUVE
observations to the predictions from the ``ORFEUS'' model and two
synthetic spectra for effective temperatures differing by $\pm 500$\,K
(corresponding to the 1\,$\sigma$ error of the Napiwotzki et al. 1995
solution). The best fit was found at $T_{\rm eff}=49\,500$\,K, in full
agreement with the ORFEUS result. Since the models with 49\,000 and
50\,000\,K already show strong deviations we believe that the EUVE
analysis improves the accuracy of the temperature determination to
about 200\,K if possible systematic errors caused by the accuracy of
the calibration are ignored.

The hydrogen column density found ($\log
N_{\rm{H}}/\rm{cm^{-2}}=18.8$) is also consistent with the value
$19.0\pm 0.4$ derived by fitting the interstellar components of the
hydrogen Lyman lines in the ORFEUS spectrum.  Therefore the value of
$\log N_{\rm{H}}/\rm{cm^{-2}} = 18.1$ determined indirectly by
Bruhweiler \& Kondo (1982) from interstellar \ion{N}{i} lines appears
clearly too low.

The overall flux distribution in both spectral ranges is well
reproduced by our LTE model. Moreover, the inconsistency between the
MW and LW range, seen in Fig.\,2 of Napiwotzki et al. (1996), is no
longer present in the newly extracted spectra.  Only in the red wing
of the \ion{He}{ii} line at 303.8\,\AA\ a small depression is visible
in the MW spectrum which is not present at the same wavelength in the
LW observation. We believe that this is due to fixed pattern noise.

The photons nominally measured at $\lambda > 456$\,\AA\ can be
entirely explained by second order contamination from the flux between
228 and 350\,\AA. Since the flux at $\lambda<228$\,\AA\ is zero, no
such contamination occurs in the MW region.

\subsubsection{Influence of hydrogen and surface gravity}
In a second step we tried to find out if our most important result
namely that about 20\%\ hydrogen is present in the atmosphere can be
confirmed by our EUVE observations. Therefore we compared the MW and
LW spectra with theoretical models for a pure helium atmosphere.
Although the flux level in the EUVE is about 30-40\%\ lower in this
case (due to a weaker Lyman absorption edge of hydrogen at 911\,\AA\
and a somewhat stronger one at 504\,\AA\ due to \ion{He}{i}) this
effect can be compensated by an increase of the effective temperature
by 200\,K and a reduction of $N_{\rm H}$ by about $2 \cdot
10^{17}$\,cm$^{-2}$, well within the error limits. If we vary our
models within the error range of hydrogen abundances from the ORFEUS
analysis the influence on the EUVE spectrum can be ignored.  This
means that the EUVE spectrum does not provide any information about
$\nhydro$ additional to the FUV region.

Likewise we have tested the influence of $\log g$ on our temperature
determination. A change of 0.08 dex corresponds to a $\approx 100$\,K
difference in $T_{\rm eff}$. In order to obtain a significant
deviation about 0.3 dex are necessary: at $\log g=7.7$ we find the
best fit at about $T_{\rm eff}=48800$\,K and $N_{\rm H}=7.4 \cdot
10^{18}$\,cm$^{-2}$, while at $\log g=8.3$ the observation is best
reproduced at $T_{\rm eff}=50\,000$\,K and $N_{\rm H}=7.8 \cdot
10^{18}$\,cm$^{-2}$.

Thus the accuracy of the temperature determination is better than
200\,K if the gravity and the hydrogen abundance are only varied
within the error limits of the ORFEUS solution.  However, the
undithered spectrum does not allow an independent simultaneous fit for
$T_{\rm eff}$, $\log g$, and $\nhydro$.

\begin{figure}[htbp]
\epsfxsize=8.8cm
\epsffile{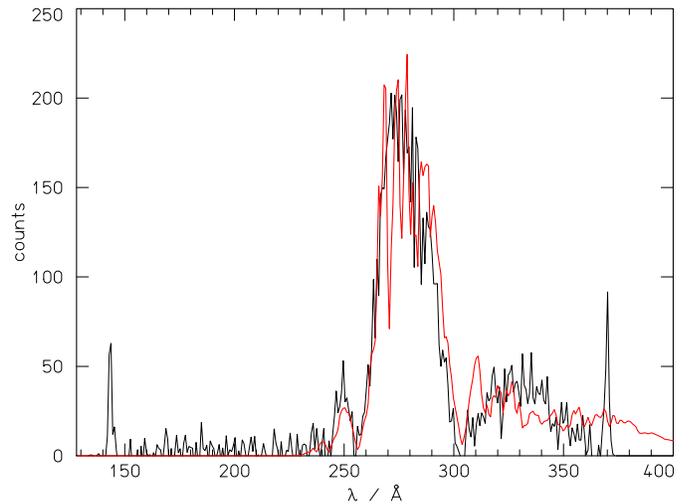}
\caption[]{A theoretical model (dotted line) with 
$T_{\rm eff}=50\,000$\,K, $\log g=8$, $\nhelium=4$ and metal
abundances as listed in Tab.\,\ref{chayer} compared to the MW spectrum
of \hdb (solid line).  $N_{\rm H}$ is assumed to be $5.8\cdot
10^{18}$\,cm$^{-2}$ }
\label{mwfity04800m1}
\end{figure}  

\begin{figure}[htbp]
\epsfxsize=8.8cm
\epsffile{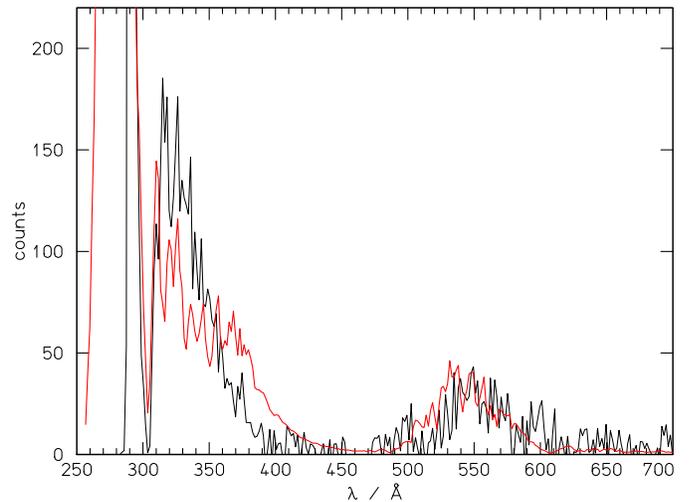}
\caption[]{The predicted LW flux for the 
same model as in Fig.\,\ref{mwfity04800m1} and the observed 
long wavelength EUVE spectrum
 }
\label{lwfity04800m1}
\end{figure}  

\subsubsection{Search for the presence of metals}
Although the EUVE spectra are contaminated by fixed pattern noise, we
have tried to find out if any of the observed features besides the
\ion{He}{ii} lines can be attributed to heavier absorbers. Metals are
expected to be present in the atmosphere due to radiative levitation:
For $T_{\rm eff}=50\,000$\,K, $\log g=8$ Chayer et al. (1995) and
Chayer (priv. comm.) predict number ratios of metals relative to
helium as listed in Tab.\,\ref{chayer}.

For this mixture and the atmospheric parameters $T_{\rm
eff}=50\,000$\,K, $\log g=8$, and $\nhydro=0.25$ we have calculated
fully line blanketed LTE model atmospheres accounting for several
million lines (taken from the Kurucz 1991 tables) by an opacity
sampling technique (resolution $\approx 0.01$\,\AA). The bound-free
opacities of C, N O, Ne, Mg, Si, S, Ar, Ca, and Fe are calculated
using Opacity Project data (Seaton 1987, Seaton et al. 1994); for Ni a
hydrogenic approximation has been used.

\begin{table}
\caption{Metal abundances as predicted by Chayer et al. (1995) for a 
DO atmosphere with $T_{\rm eff}=50\,000$\,K and  $\log g=8$}
\begin{tabular}{r@{\,\,=\,\,}l}
\hline
\multicolumn{2}{c}{$\log$ number ratio} \\
\hline
  C/He &$-4.10$\\
  N/He &$-4.03$\\
  O/He &$-4.20$\\
 Ne/He &$-4.10$\\
 Na/He &$-5.39$\\
\hbox{}\hspace{2.5mm} Mg/He & $-\infty$\\
 Al/He & $-\infty$\\
 Si/He & $-\infty$\\
  S/He &$-4.67$\\
 Ar/He &$-4.20$\\
 Ca/He &$-4.44$\\
 Fe/He &$-4.52$\\
\hline
\end{tabular}
\label{chayer}
\end{table}

Theoretical EUVE spectra for this composition are shown in
Fig.\,\ref{mwfity04800m1} and \ref{lwfity04800m1}: in both, the MW and
the LW range, strong features occur, some of them being in clear
disagreement with the observation. Moreover, the fit to the overall
energy distribution becomes somewhat worse compared to
Fig.\,\ref{xmwfit04800} and Fig.\,\ref{xlwfit04800} where atmospheres
containing only hydrogen and helium have been used.

Finally, we tried to find out if any of the features in the EUVE
spectra can be attributed to individual elements. We calculated
theoretical spectra for $T_{\rm eff}=50\,000$\,K, $\log g=8$,
$\nhydro=0.25$ and a single additional element with an abundance given
by Table\,\ref{chayer}. The results are:
\begin{itemize}
\item
 None of the features predicted forcarbon can be identified in the SW
or MW spectra. The number ratio of $\log {\rm C/He} = -4.10$ can be
regarded as an upper limit since some of the absorption lines are
already stronger than the ``noise'' level.
\item
 A nitrogen abundance of $\log {\rm N/He}= -4.03$ can be definitely
excluded due to a rather strong \ion{N}{iv} absorption feature at
about 283\,\AA\ which is not observed.
\item
At about the same wavelength the only significant absorption line of
Ne occurs. At $\log {\rm Ne/He}= -4.1$ it is, however, much weaker so
that Ne can in principle be present even at a somewhat higher
abundance.
\item
Sodium does not exhibit any strong features in the EUVE so that its
presence can neither be confirmed nor excluded.
\item
Due to a relatively strong predicted \ion{S}{iii} line at 276\,\AA\ we
can conclude that the number ratio cannot be much higher than $\log
{\rm S/He} = -4.67$.
\item
Likewise a feature at 339\,\AA\ in the theoretical spectrum is still
compatible with the observation if $\log {\rm Ar/He} = -4.20$ is
assumed but can also be regarded as an approximate upper limit.
\item
From the mixture given in Table.\,\ref{chayer} the strongest
absorption lines are predicted for Ca and Fe.  Both, $\log {\rm Ca/He}
= -4.44$ and $\log {\rm Fe/He} = -4.52$, are clearly at variance with
the observation and must be lower by at least one dex.
\end{itemize}

For a more precise analysis a dithered EUVE spectrum would be
necessary.

\section{Discussion}
The \hd\ system is a bright source in the PSPC camera of ROSAT. We
showed that the measured X-ray flux is attributed to the main sequence
component HD149499A. It indicates a high level of coronal activity in
this K0V star.

\hdb\ is a rather unique object since it is one of the coolest DO 
stars, close to the DB gap, and it is the only DO in which hydrogen
 has been detected (Napiwotzki et al.\ 1995).

Our goal was to investigate if the MW and LW EUVE spectra of \hdb\ can
further constrain the atmospheric parameters derived by our analysis
of the ORFEUS spectrum. Since only two spectral lines of \ion{He}{ii}
are visible in the EUVE spectra well above the noise level, we could
not use the spectra for a simultaneous determination of $T_{\rm eff}$,
$\log g$, and $\nhydro$, and the interstellar column density. However,
the EUVE observation is consistent with the ORFEUS result: For $\log
g=7.97\pm 0.08$ and $\nhydro=0.22\pm 0.11$ we obtain $T_{\rm
eff}=49\,500$\,K with an accuracy of about 200\,K. A precise
temperature determination is rather important since currently the blue
edge of the DB gap is not well determined; there is no known DO white
dwarf significantly cooler than 50\,000\,K.

Unfortunately, the analysis of the EUVE data does not set any further
limits to the hydrogen abundance and the gravity, since  changes
in $\nhydro$ or $\log g$ can be compensated by rather small changes in
$T_{\rm eff}$ and $N_{\rm H}$.

With the help of fully line blanketed LTE model atmospheres we could
show that the the mixture of metals predicted by the current diffusion
theory (Chayer et al.\ 1995) is in worse agreement with the EUVE
observations than a H+He model.  Although no individual elements could
be identified in the spectra we conclude that iron, calcium, and
nitrogen are less abundant compared to the Chayer et al. values;
moreover, the content of carbon, sulphur, and argon in the atmosphere
of \hdb\ is at least not higher than predicted. Carbon is the only
other element beside helium which could be identified by high
resolution UV spectroscopy.  Werner \&\ Dreizler (1996) derived
$n_{\rm C}/n_{\rm He}=10^{-5}$ from IUE data.

While pratically all DA white dwarfs are metal deficient relative to
the predictions of the diffusion theory (Jordan et al. 1996a, 1996b;
Finley 1996), Dreizler \&\ Werner (1996) have shown that among the DO
stars both cases exist, those where the observed abundances of C, N,
O, and Si are higher than the predictions and those where the number
ratios are lower.  These confusing results show that it is necessary
to improve the theory of diffusion in hot white dwarfs.

\acknowledgements
This research has made use of the SIMBAD database, operated at CDS,
Strasbourg, France. Work on ROSAT and ORFEUS in Kiel was supported by
DARA grant 50 OR 94091.


\begin{thebibliography}{}
\bibitem{}
 Barstow, M.A., Holberg J.B., Koester, D., Nousek J.A., Werner, K.,
1995, in White Dwarfs, Lecture Notes in Physics, eds. D. Koester \&\
K. Werner, Springer, p. 302
\bibitem{}  Bruhweiler, F.C., Kondo, Y. ,1982, \apj\ 259, 232
\bibitem{} Chayer, P., Fontaine, G., Wesemael, F., 1995, ApJS 99, 189
\bibitem{} Dreizler, S., Werner, K., 1996, A\&A, in press
\bibitem{}  Feibelman, W.A., Bruhweiler, F.C., 1990, \apj\ 357, 548
\bibitem{}  Finley D.S., 1996, in Astrophysics in the Extreme Ultraviolet,
eds. S. Bowyer and R.F. Malina, Kluwer, p. 223
\bibitem{} Hempelmann, A., Schmitt, J.H.M.M., Schultz, M., R\"udiger, G., 
St\c{e}pie{\'n}, K. 1995, A\&A 294, 515
\bibitem{} Ianna, P.A., Rohde, J.R., Newell, E.B., 1982, \apj\ 259, L71
\bibitem{}
Jordan, S., Wolff, B., Koester, D., Napiwotzki, R. 1994, A\&A  290, 834
\bibitem{}  Jordan, S., Koester, D., Finley, D., 1996a,
in Astrophysics in the Extreme Ultraviolet,
eds. S. Bowyer and R.F. Malina, Kluwer, p. 235
\bibitem{}
Jordan, S., Finley, D.,  Koester, D.,  Wolff, B., 1996b,
in R\"ont\-gen\-strah\-lung from the Universe, MPE Report 263, p. 5
\bibitem{}
 Koester, D., Liebert, J., Hege, E.K., 1979, \aua\ 71, 163
\bibitem{}  Kurucz R.L., 1991, in Stellar Atmospheres: Beyond Classical
Models, NATO ASI Series, 341, p.441
\bibitem{} Liebert, J., Wesemael, F., Hansen, C.J. et al., 1986, ApJ 309, 241
\bibitem{}Morrison R., MacCammon D., 1983  ApJ  270, 119
\bibitem{}
Napiwotzki, R., Hurwitz, M.,  Jordan, S.,  Bowyer, S.,  Koester, D.,
 Weidemann, V., Lampton, M., Edelstein J., 1995, A\&A 300, L5
\bibitem{}
Napiwotzki, R., Jordan, S., Bowyer, S., Hurwitz, M.,  Koester, D.,
Rauch, T.,  Weidemann, V., 1996, in Astrophysics in the Extreme Ultraviolet,
eds. S. Bowyer \& R.F. Malina, Kluwer, p. 241
\bibitem{} Pye J.P., McGale P.A., Allan D.J., et al., 1995, MNRAS 274, 1165
\bibitem{} Seaton, M.J., 1987, JPhysB 20, 6363
\bibitem{} Seaton, M.J., Yan Y., Mihalas, D., Pradhan, A. K.,
1994, MNRAS 266, 805
\bibitem{}Rumph T., Bowyer S., Vennes S., 1994, AJ  107, 2108
\bibitem{}
 Sion, E.M., Guinan, F., Wesemael, F., 1982, \apj\ 255, 232
\bibitem{} Vennes, S.,  Chayer, P.,  Fontaine, G., Wesemael, F., 1989, 
ApJ 336, L25
\bibitem{}
Werner, K. 1986, \aua\ 161, 177
\bibitem{}
Werner, K. 1993, in: White dwarfs: advances in observation and theory, ed.\
	M.A.~Barstow, Kluwer, Dordrecht, p.~67
\bibitem{}
Werner, K., Dreizler, S. , Wolff, B. 1995, A\&A 298, 567
\bibitem{}
 Wray, J.D., Parsons, S.B., Henize, K.G., 1979, \apj\ 234, L187
\end{thebibliography}
\end{document}